\documentclass[12pt]{article}
\usepackage{graphicx}

\newcommand\pubnumber{}
\newcommand\pubdate{\today}

\def\cernois{CERN, 1211 Geneva 23, Switzerland}
\def\Title#1{\begin{center} {\Large #1 } \end{center}}
\def\Author#1{\begin{center}{ \sc #1} \end{center}}
\def\Address#1{\begin{center}{ \it #1} \end{center}}

\newcommand\pubblock{\rightline{\begin{tabular}{l} \pubnumber\\
         \pubdate  \end{tabular}}}
\newenvironment{Abstract}{\begin{quotation}  }{\end{quotation}}
\newenvironment{Presented}{\begin{quotation} \begin{center} 
             PRESENTED AT\end{center}\bigskip 
      \begin{center}\begin{large}}{\end{large}\end{center} \end{quotation}}

\usepackage{amssymb}
\usepackage{xspace}

\def\beq{\begin{equation}}
\def\eeq#1{\label{#1}\end{equation}}
\def\eeqn{\end{equation}}

\def\beqa{\begin{eqnarray}}
\def\eeqa#1{\label{#1}\end{eqnarray}}
\def\eeqan{\end{eqnarray}}

\let\bar=\overbar

\def\ie{{\it i.e.}}
\def\eg{{\it e.g.}}

\def\D{{\cal D}}

\def\Dslash{\not{\hbox{\kern-4pt $D$}}}
\def\dslash{\not{\hbox{\kern-2pt $\del$}}}

\def\msb{{\bar{\ssstyle M \kern -1pt S}}}

\def\DDbar   {\ensuremath{\kern -0.1em \stackrel{\kern 0.1em \textsf{\fontsize{5pt}{1em}\selectfont(---)}}{D}\kern -0.3em}\xspace}
\def\AAbar   {\ensuremath{\kern -0.2em \stackrel{\kern 0.2em \textsf{\fontsize{5pt}{1em}\selectfont(---)}}{A}\kern -0.3em}\xspace}
\def\AfAfbar   {\ensuremath{\kern -0.2em \stackrel{\kern 0.2em \textsf{\fontsize{5pt}{1em}\selectfont(---)}}{A}_{\kern -0.3em f}\kern -0.3em}\xspace}
\def\dacp     {\ensuremath{\Delta A_{\CP}}\xspace}
\def\CP                {\ensuremath{C\!P}\xspace}
\def\Dbar    {\kern 0.2em\overline{\kern -0.2em \PD}{}\xspace}

\def\Dz      {\ensuremath{\PD^0}\xspace}
\def\Dzb     {\ensuremath{\Dbar^0}\xspace}
\def\DzDzb   {\ensuremath{\Dz {\kern -0.16em \Dzb}}\xspace}
\def\Dp      {\ensuremath{\PD^+}\xspace}
\def\Dm      {\ensuremath{\PD^-}\xspace}
\def\Dpm     {\ensuremath{\PD^\pm}\xspace}

\def\DpDm    {\ensuremath{\Dp {\kern -0.16em \Dm}}\xspace}

\def\Dstarp  {\ensuremath{\PD^{*+}}\xspace}

\def\mup      {\ensuremath{\mu^+}\xspace}
\def\mum      {\ensuremath{\mu^-}\xspace}

\def\lhcb {LHCb\xspace}

\def\PD      {\ensuremath{D}\xspace}                 
\def\PM      {\ensuremath{M}\xspace}                 
\def\PB      {\ensuremath{B}\xspace}

\def\PK      {\ensuremath{K}\xspace}                 
\def\kaon  {\ensuremath{\PK}\xspace}
\def\Kp    {\ensuremath{\kaon^+}\xspace}
\def\Km    {\ensuremath{\kaon^-}\xspace}

\def\Kmp   {\ensuremath{\kaon^\mp}\xspace}
\def\pipm   {\ensuremath{\Ppi^\pm}\xspace}

\def\Ppi         {\ensuremath{\pi}\xspace}                 
                 
\def\pion  {\ensuremath{\Ppi}\xspace}
\def\pip   {\ensuremath{\pion^+}\xspace}
\def\pim   {\ensuremath{\pion^-}\xspace}
\def\ycp        {\ensuremath{y_{CP}}\xspace}
\def\agamma     {\ensuremath{A_{\Gamma}}\xspace}
\def\KS    {\ensuremath{\kaon^0_{\rm\scriptstyle S}}\xspace} 
 
\def\Bbar    {\kern 0.2em\overline{\kern -0.2em \PB}{}\xspace}

\def\squark    {\ensuremath{\Ps}\xspace}
\def\Ps      {\ensuremath{\mathrm{s}}\xspace}

\newcommand{\decay}[2]{\ensuremath{#1\!\to #2}\xspace}         % {\Pa}{\Pb \Pc}
\def\DDbar   {\ensuremath{\kern -0.1em \stackrel{\kern 0.1em \textsf{\fontsize{5pt}{1em}\selectfont(---)}}{D}\kern -0.3em}\xspace}
\def\AAbar   {\ensuremath{\kern -0.2em \stackrel{\kern 0.2em \textsf{\fontsize{5pt}{1em}\selectfont(---)}}{A}\kern -0.3em}\xspace}
\def\AfAfbar   {\ensuremath{\kern -0.2em \stackrel{\kern 0.2em \textsf{\fontsize{5pt}{1em}\selectfont(---)}}{A}_{\kern -0.3em f}\kern -0.3em}\xspace}
\def\dacp     {\ensuremath{\Delta A_{\CP}}\xspace}

\def\Mbar    {\kern 0.2em\overline{\kern -0.2em \PM}{}\xspace}

\def\mup        {\ensuremath{\Pmu^+}\xspace} % muon negative (\mum is taken)
\def\Pmu         {\ensuremath{\mu}\xspace}

\def\mup        {\ensuremath{\Pmu^+}\xspace}
\def\pion  {\ensuremath{\Ppi}\xspace}

\def\pip   {\ensuremath{\pion^+}\xspace}
\def\pim   {\ensuremath{\pion^-}\xspace}
\def\lhcb {LHCb\xspace}
\def\ux85 {UX85\xspace}

  \def\Dbar    {\kern 0.2em\overline{\kern -0.2em \PD}{}\xspace}
\def\D       {\ensuremath{\PD}\xspace}

\def\Dz      {\ensuremath{\D^0}\xspace}
\def\Dzb     {\ensuremath{\Dbar^0}\xspace}
\def\DzDzb   {\ensuremath{\Dz {\kern -0.16em \Dzb}}\xspace}
\def\Dp      {\ensuremath{\D^+}\xspace}
\def\Dm      {\ensuremath{\D^-}\xspace}
\def\Dpm     {\ensuremath{\D^\pm}\xspace}

\def\DpDm    {\ensuremath{\Dp {\kern -0.16em \Dm}}\xspace}

\def\Dstarp  {\ensuremath{\D^{*+}}\xspace}

\def\Dsp     {\ensuremath{\D^+_\squark}\xspace}

\def\invfb   {\ensuremath{\mbox{\,fb}^{-1}}\xspace}
\def\ns   {\ensuremath{\mbox{\,ns}}\xspace}
\def\mhz  {\ensuremath{\mbox{\,MHz}}\xspace}
\def\khz  {\ensuremath{\mbox{\,kHz}}\xspace}

\newcommand{\tev}{\ensuremath{\mathrm{\,Te\kern -0.1em V}}\xspace}
\newcommand{\gev}{\ensuremath{\mathrm{\,Ge\kern -0.1em V}}\xspace}
\newcommand{\mev}{\ensuremath{\mathrm{\,Me\kern -0.1em V}}\xspace}
\newcommand{\kev}{\ensuremath{\mathrm{\,ke\kern -0.1em V}}\xspace}
\newcommand{\ev}{\ensuremath{\mathrm{\,e\kern -0.1em V}}\xspace}
\newcommand{\gevc}{\ensuremath{{\mathrm{\,Ge\kern -0.1em V\!/}c}}\xspace}
\newcommand{\mevc}{\ensuremath{{\mathrm{\,Me\kern -0.1em V\!/}c}}\xspace}
\newcommand{\gevcc}{\ensuremath{{\mathrm{\,Ge\kern -0.1em V\!/}c^2}}\xspace}
\newcommand{\gevgevcccc}{\ensuremath{{\mathrm{\,Ge\kern -0.1em V^2\!/}c^4}}\xspace}
\newcommand{\mevcc}{\ensuremath{{\mathrm{\,Me\kern -0.1em V\!/}c^2}}\xspace}

\begin{document}
\begin{titlepage}
\pubblock

\vfill
\Title{The charm physics programme at the LHCb upgrade, and Atlas and CMS upgrades}
\vfill
\Author{Marco Gersabeck\\ on behalf of the LHCb collaboration}
\Address{\cernois}
\vfill
\begin{Abstract}
Charm physics has been established at the LHC based on several high-precision measurements.
The future of charm physics at the LHC experiments is discussed in detail.
The bulk of the charm physics programme will be performed by LHCb and the LHCb upgrade.
In particular, the impact of the LHCb upgrade on mixing and \CP violation measurements is presented.
\end{Abstract}
\vfill
\begin{Presented}
The $5^{th}$ International Workshop on Charm Physics\\
Honolulu, Hawai'i, USA,  14--17 May 2012
\end{Presented}
\vfill
\end{titlepage}
\def\thefootnote{\fnsymbol{footnote}}
\setcounter{footnote}{0}

\section{Introduction}
The LHC has performed excellently in its first years of operation and provided large datasets at unprecedented collision energies.
With the data recorded at the four interaction points all experiments have proven their feasibility and outstanding operational quality.
Charm physics has been an integral part of this road to success, ranging from first cross-section measurements at $\sqrt{s}=7\tev$~\cite{LHCb-CONF-2010-013} to the first evidence of \CP violation in the charm system~\cite{Aaij:2011in}.
A full account of the current status of charm physics is given in Reference~\cite{Gersabeck:2012rp}.

\section{The LHC upgrade schedule}
The first running phase of the LHC will last until the end of 2012 for proton-proton collisions, followed by a roughly four-week run of proton-lead collisions at the beginning of 2013.
This is followed by an approximately 18-months shutdown (LS1) for maintenance and consolidation work after which the LHC is expected to operate close to its design collision energy of $13$--$14\tev$.
In addition, also the filling scheme is expected to go from the current $50\ns$ bunch spacing to the nominal $25\ns$ bunch spacing, \ie\ doubling the number of bunches.
This break also allows first work on detector upgrade installations.
The second running period is foreseen to last from late 2014 until 2017 included, followed by another long shutdown (LS2) which will see significant work on detector upgrades.
The running period beyond 2018 will head towards the high-intensity LHC, following another long shutdown around 2022 (LS3).

\section{The detector upgrade plans}
The Atlas collaboration plans to pursue their charm programme until LS2.
The main relevant detector upgrade is thus the installation of a fourth barrel pixel layer in LS1.
With an improved impact parameter resolution this is of particular importance to the heavy flavour physics programme.
The CMS collaboration has no plans to pursue a charm programme after LS1.

LHCb, as a dedicated heavy flavour experiment, will continue its charm programme throughout its running period.
A major upgrade of the experiment is foreseen to be installed in LS2.
The LS1 will be used to adapt the existing data acquisition and processing infrastructure to the changes in accelerator conditions after the shutdown.
The key challenge when operating at higher instantaneous luminosities is the efficiency for collecting data of hadronic decays.
Based on the current LHCb layout, the hadronic trigger efficiency reduces due to harsher cuts in the hardware trigger stage.
This leads to a saturation of the signal yield as function of instantaneous luminosity.

The solution to allow operation at higher luminosities is a complete re-design of the trigger system.
The LHCb upgrade is based on the ability to read out the detector at the LHC clock frequency of $40\mhz$.
The current first trigger level, which reduced the data rate to $1\mhz$, will be replaced by a flexible custom-electronics trigger which can be tuned to output rates between the current $1\mhz$ and the full $40\mhz$.
This is followed by a software-based trigger stage which, using the full detector information, has to reduce the output rate to $20\khz$, \ie\ a factor of four higher than the current output rate.

The upgrade task of the sub-detector system is to maintain the current high performance while providing the possibility of the $40\mhz$ readout in the presence of a significantly increased particle rate.
The LHCb upgrade is currently in the design phase and TDR documents which finalise several technology choices are planned for 2013.
A framework TDR outlining the various detector options as well as the financial planning and insitute's interests has been submitted to LHCC in June 2012~\cite{FTDR}.
The vertex detector will be replaced either by a silicon-strip detector with finer strip pitch or by a silicon-pixel detector of similar layout.
The remaining tracking system will be replaced by a combination of silicon detectors and possibly scintillating fibre or straw detectors.
The RICH detectors have to be equipped with new photon detectors which can be read out at the increased rate.
The calorimeters have to receive new readout electronics. 

The data yield per year is expected to increase for several reasons.
The charm production cross-section is projected to increase by a factor of about $1.8$ when going to nominal LHC energy.
This gain will take effect already after LS1.
The trigger efficiency is assumed to increase by a factor of $2$, however, this factor may be significantly larger for multi-body decay modes.
With the annual integrated luminosity expected to increase by a factor between $3$ and $5$, the annual signal yield is estimated to increase by about an order of magnitude.
The total integrated luminosity recorded during the LHCb upgrade period is assumed to be $50\invfb$.
As an example, this leads to an expectation of $4\times10^{10}$ offline selected \decay{\Dz}{\Km\pip} decays.

\section{Charm production and spectroscopy}
The LHCb collaboration has recently published a set of measurements of the production of double-charm events, \ie\ events containing double charmonium, charmonium plus open charm, or double open charm~\cite{Aaij:2012dz}.
These studies will be continued with increased data samples.
They will be extended by studies of the production of other charmonium modes, both at Atlas and at LHCb.
This will be complemented by studies of the combined production of charmonium and jets or vector bosons.

Another topic requiring high luminosity and hence the upgrade is the search for doubly-heavy and triply-heavy baryons.
Given their production cross-sections which fall sharply with increased transverse momentum~\cite{Chen:2011mb} these need large data samples to be discovered.
At the same time, it is mandatory to have an efficient triggering to reconstruct these complex multi-body final states without requiring large transverse momenta.
Beyond the discovery of new states it is of interest to study their properties such as lifetimes, branching ratios, quantum numbers, and their spectrum of excited states.

\section{Rare decay searches and analyses}
Rare decay searches naturally gain in sensitivity with increasing luminosity and hence benefit from the increased data sets planned for the detector upgrades at the LHC.
A second important component to maximise sensitivity is a low level of background.
For this reason only LHCb plans to continue rare decay searches with their upgraded detector.

%%%%%%%%%%%%%%%%%%%%%%%%%%%%%%%%%%%%%%%%%%%%%%%%%%%%%%%%%%%%%%%%%%%%%%%%
\begin{figure}[htb]
\centering
\includegraphics[width=\textwidth]{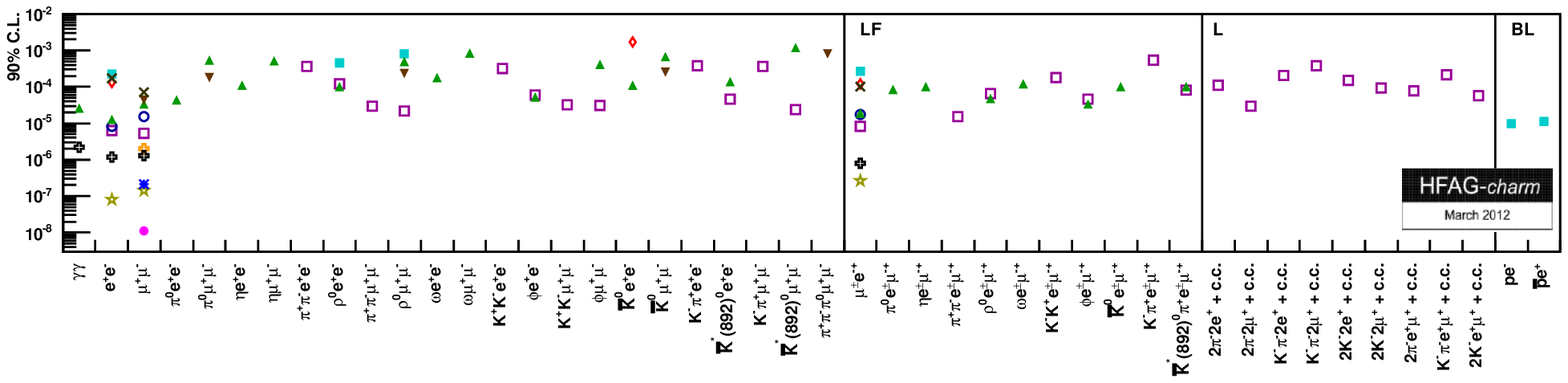}\\
\includegraphics[width=0.3\textwidth]{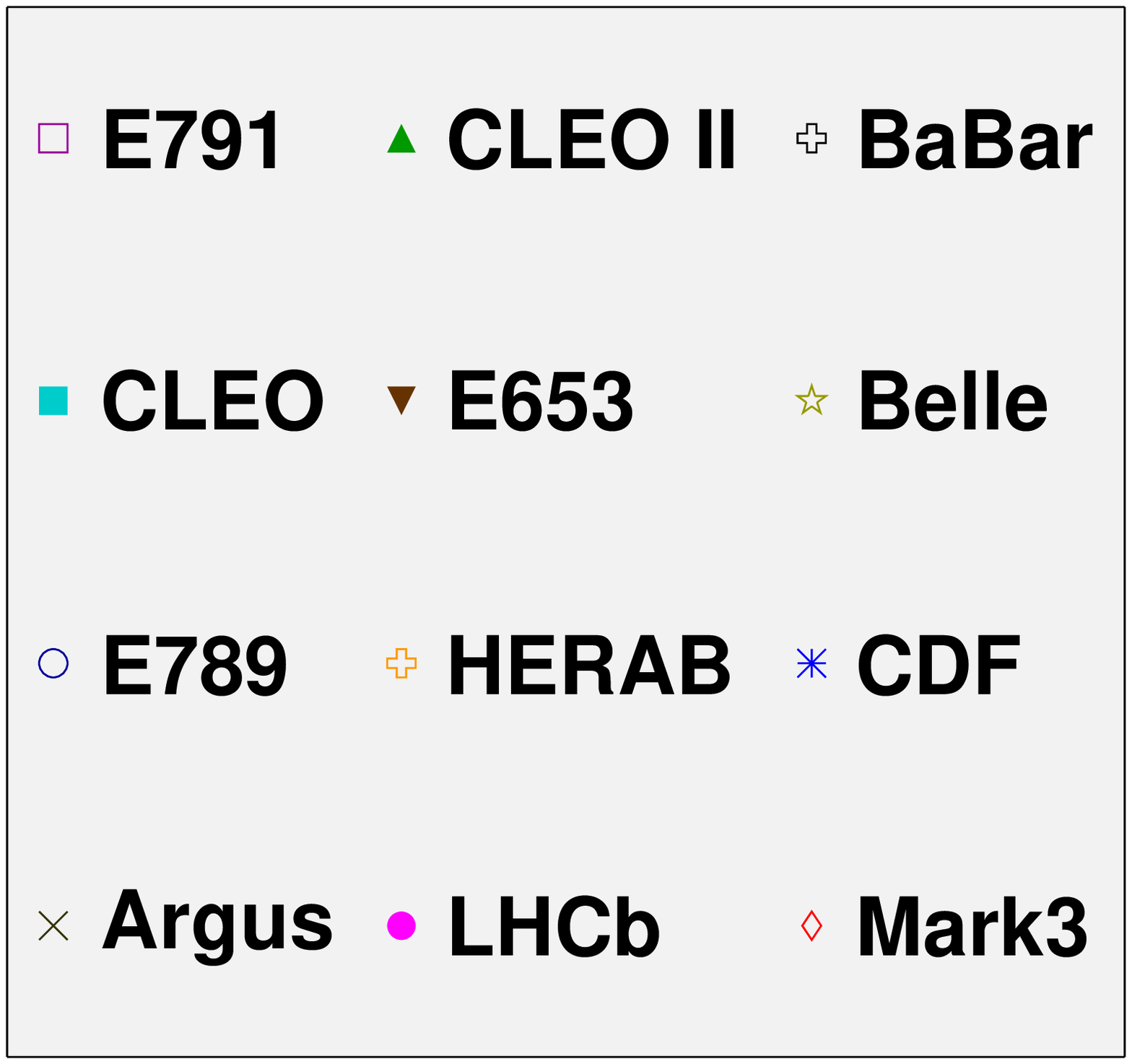}
\caption{Current best limits on \Dz decays. The different regions indicate flavour changing neutral current decays, lepton-flavour violating decays (LF), lepton-number violating decays (L), and lepton and baryon-number violating decays (BL). Reproduced from Ref.~\cite{Amhis:2012hf}.}
\label{fig:rare_d0}
\end{figure}
%%%%%%%%%%%%%%%%%%%%%%%%%%%%%%%%%%%%%%%%%%%%%%%%%%%%%%%%%%%%%%%%%%%%%%%%%%

LHCb recently set the best limit for the flavour-changing neutral-current decay \decay{\Dz}{\mum\mup} at $1.1\times10^{-8}$~\cite{LHCb-CONF-2012-005} (see Figure~\ref{fig:rare_d0}).
With the LHCb upgrade this limit is expected to be improved by about one order of magnitude.

Another set of promising measurements are those of the dimuon invariant-mass spectrum of \decay{\Dpm}{h^\pm\mum\mup} decays.
The LHCb acceptance allows the study of the full spectrum down to the kinematic threshold to search for suppressed resonances such as sGolstinos.
The same decay with a same-sign lepton pair is one example for a lepton-number violating decay.
This decay can be mediated for example by Majorana neutrinos.

Four-body decays offer several different approaches for searching for physics beyond the standard model.
An interesting example among rare decays is the decay \decay{\Dz}{\Km\Kp\mum\mup}.
Beyond the search and eventual discovery of this mode it is of interest to study its symmetries.
The measurement of the forward-backward asymmetry of one of the muons, the search for \CP violation, and the measurement of T-odd correlations provide complementary information~\cite{Bigi:2009jj}.
Assuming a branching fraction of $10^{-6}$ the LHCb upgrade should provide a sample of several hundred selected candidates.
In comparison, the decay \decay{\Dz}{\Km\Kp\pim\pip} is expected to yield a sensitivity to T-odd correlations of $2.5\times10^{-4}$ at the end of the LHCb upgrade.

\section{Mixing measurements}
There are a number of open questions in the area of charm mixing and \CP violation.
The mixing mechanism is well established through a number of complementary measurements.
However, there is still no precise determination of the underlying parameters.
The LHCb collaboration has reported first evidence for \CP violation in the charm system~\cite{LHCb-PAPER-2011-023}, but it is not clear whether this is due to physics beyond the standard model or whether it can be explained by hadronic uncertainties within the standard model.
Beyond this measurement there is no further evidence for \CP violation in the charm sector to date.
In particular, indirect \CP violation is still out of reach.

The ultimate goal of mixing and \CP violation measurements in the charm sector is to reach the precision necessary to probe the standard model predictions or better.
Theoretically, precise standard model predictions are still challenging.
As a general goal, $10\%$ relative uncertainties on the underlying parameters should be achieved.
There are a number of possible ways to measure mixing parameters.
Possibly the most powerful is the observable \ycp, measured as a ratio of effective lifetimes in the \CP eigenstates $\Km\Kp$ or $\pim\pip$ with respect to the Cabibbo-favoured state $\Kmp\pipm$.
The recent LHCb measurement, based on the small data sample recorded in 2010, is a proof of principle for such measurements at a hadron collider.
Its systematic uncertainty is expected to scale well with the increasing sample size as more sophisticated treatments of the contributing backgrounds become possible.

Another mixing measurement based on two-body decays is that using the wrong-sign decay \decay{\Dz}{\Kp\pim}.
This measurement requires external information on the strong phase shift to yield information on the mixing parameters themselves.
The measurement of the forbidden decay rate \decay{\Dz}{\Kp\mum\nu} which is only accessible through mixing gives access to the absolute mixing rate $R_m\equiv(x^2+y^2)/2$.
Through a time-dependent Dalitz analysis, the decays \decay{\Dz}{\KS\pim\pip} and \decay{\Dz}{\KS\Km\Kp} give access to the individual mixing parameters $x$ and $y$.
The expected sensitivities are given in Table~\ref{tab:mixing}.

%%%%%%%%%%%%%%%%%%%%%%%%%%%%%%%%%%%%%%%%%%%%%%%%%%%%%%%%%%%%%%%%%%%%%%%%%
\begin{table}[hbt]
\begin{center}
\begin{tabular}{l|ccc}  
Decay & Observable & Expected sensitivity (in $10^{-3}$) \\ \hline
 \decay{\Dz}{\Km\Kp} &   \ycp     &    $0.04$   \\
 \decay{\Dz}{\pim\pip} &   \ycp     &    $0.08$   \\
 \decay{\Dz}{\Kp\pim} &  $x'^2$,~$y'$     &    $0.01$,~$0.1$   \\ 
 \decay{\Dz}{\KS\pip\pim} &  $x$,~$y$     &    $0.15$,~$0.1$   \\ 
 \decay{\Dz}{\Kp\mum\nu} &  $x^2+y^2$     &    $0.0001$   \\ \hline
\end{tabular}
\caption{Expected statistical sensitivities for mixing observables for $50\invfb$.}
\label{tab:mixing}
\end{center}
\end{table}
%%%%%%%%%%%%%%%%%%%%%%%%%%%%%%%%%%%%%%%%%%%%%%%%%%%%%%%%%%%%%%%%%%%%%%%%%%%

\section{\CP violation measurements}
\subsection{Indirect \CP violation}
Indirect \CP violation measurements at \lhcb are mostly constrained by the observable \agamma~\cite{Gersabeck:2011xj}, which is the asymmetry of effective lifetimes of \Dzb and \Dz decays to a \CP eigenstate and which can be written as
\begin{equation}
\agamma\approx\frac{1}{2}(A_m+A_d)y\cos\phi-x\sin\phi\approx-a_{\CP}^{\rm ind}-a_{\CP}^{\rm dir}\ycp,
\end{equation}
where $A_m$ is the deviation of $|q/p|$ from $1$, $A_d$ is the deviation of $|A_f/\overline{A}_f|$ from $1$, and $\phi$ is the relative weak phase between the two fractions, following Reference~\cite{Gersabeck:2012rp}.
The \CP violating parameters in this observable are multiplied by the mixing parameters $x$ and $y$, respectively.
Hence, the relative precision on the \CP violating parameters is limited by the relative precision of the mixing parameters.
Therefore, aiming at a relative precision below $10\%$ and taking into account the current mixing parameter world averages, the target precision for the mixing parameters would be $2-3\times10^{-4}$, which can be reached with the LHCb upgrade.
With standard model indirect \CP violation expected to be of the order of $10^{-4}$, the direct \CP violation parameter contributing to \agamma has to be measured to a precision of $10^{-3}$ in order to distinguish the two types of \CP violation in \agamma.
Based in the existing LHCb measurement~\cite{Aaij:2011ad}, the ultimate statistical precision expected for \agamma is better than $10^{-4}$ in both decays to \Km\Kp and to \pim\pip.

The above mentioned decays \decay{\Dz}{\KS\pim\pip} and \decay{\Dz}{\KS\Km\Kp} give access not only to the mixing parameters but also to the \CP-violating parameters $|q/p|$ and $\phi$.
Two different approaches exist to extract these parameters: a model-dependent measurement of the time evolution of Dalitz-plot parameters, and a model-independent measurement of the time evolution in regions of different strong-phase difference.
The latter approach requires input from measurements using quantum-correlated charm meson pairs produced at threshold.
Such measurements exist from CLEOc (see \eg~\cite{Libby:2010nu}) and can be performed by the BESIII collaboration in the future.
It is important that the existing measurements are further improved in precision to minimise systematic limitations in this approach.

The interplay between indirect and direct \CP-violation parameters shows the importance of measuring both sets of parameters in order to interpret the result.
At the same time precise theory predictions are required to identify the source of \CP violation.
Where this is not possible it may be feasible to constrain theoretical uncertainties through theoretically clean control measurements.
Beyond complementary \CP violation measurements such control measurements can also be ratios of particle lifetimes~\cite{Bobrowski:2012jf}.
LHCb has shown that lifetime ratio measurements involving different hadronic final states are feasible at hadron colliders~\cite{Aaij:2011ad}.

\subsection{Direct \CP violation}
Measurements of direct \CP violation at the LHC are linked to several challenges.
They require control of the production asymmetries present in the proton-proton collisions.
Furthermore, several sources of detection asymmetries can arise.
When measuring a \Dz decay the flavour at production is commonly determined by the charge of the pion from a \Dstarp decay which is subject to detection asymmetries.
The usage of \Dz mesons originating in semileptonic \PB decays, where the lepton charge determines the flavour, is another possibility and subject to similar detection asymmetries.
While some decays have pairs of oppositely charged hadrons of the same type, for which detection asymmetries cancel provided sufficient kinematic overlap, charged \PD decays and some \Dz decays will have unmatched decay products leading to inevitable detection efficiency effects.

If independent measurements of the asymmetries masking the \CP asymmetry are not available it is helpful to construct new observables where some asymmetries cancel.
Differences of asymmetries are a common choice as the measured asymmetries can be expressed as sums of individual asymmetries to first order, provided all asymmetries are small.
Differences of similar final states, \eg\ two singly Cabibbo-suppressed decays, have the advantage that potentially all unwanted asymmetries cancel, but the measured asymmetry is the difference of two \CP asymmetries.
Access to individual asymmetries can be obtained by using a Cabibbo-favoured decay in the difference and assuming no \CP violation in this decay.
While this cancels the production asymmetry it usually leaves a detection asymmetry due to the difference in the final states.
The comparison of different Cabibbo-favoured decays can give access to detection asymmetries.

At the levels of precision anticipated for the LHCb upgrade control of these asymmetries will be paramount for measuring time-integrated \CP asymmetries.
In addition to the challenges of measuring asymmetries to sub-percent level precision, the non-cancellation due to second-order effects in asymmetries will have to be taken into account.
In searches for \CP violation in the phase space of a multi-body decay production asymmetries largely drop out as they are constant throughout the phase space.
Detection asymmetries may vary more strongly as they depend on the daughter particle momentum which naturally varies across phase space.
However, a signal for \CP violation in a narrow resonance should still be distinguishable from a detection efficiency effect.
The following measurements are examples for those foreseen during the upgrade period.

Table~\ref{tab:cpvdir} summarises several sensitivity estimates for \CP violation measurements.
For two-body \Dz decays, the individual \CP asymmetries will be measured as well, however, their final precision critically depends on the knowledge of production and detection asymmetries as described above.
In addition to the two-body and three-body \Dp decays, similar decay modes of \Dsp will be analysed.
For multi-body decays several types of model-independent searches will be exploited as well as model dependent studies.

%%%%%%%%%%%%%%%%%%%%%%%%%%%%%%%%%%%%%%%%%%%%%%%%%%%%%%%%%%%%%%%%%%%%%%%%%
\begin{table}[hbt]
\begin{center}
\begin{tabular}{l|ccc}  
Decay & Observable & Expected sensitivity (in $10^{-3}$) \\ \hline
 \decay{\Dz}{\Km\Kp},\decay{\Dz}{\pim\pip} &   \dacp     &    $0.15$   \\
 \decay{\Dp}{\KS\Kp} &   $A_{\CP}$     &    $0.1$   \\
 \decay{\Dp}{\Km\Kp\pip} &   $A_{\CP}$    &    $0.05$   \\ 
 \decay{\Dp}{\pim\pip\pip} &  $A_{\CP}$     &    $0.08$   \\ 
 \decay{\Dp}{h^-h^+\pip} & CPV in phases  &    $(0.01-0.1)^\circ$ \\
 \decay{\Dp}{h^-h^+\pip} & CPV in fractions &    $0.1-1.0$\\\hline
\end{tabular}
\caption{Expected statistical sensitivities for direct \CP violation observables for $50\invfb$.}
\label{tab:cpvdir}
\end{center}
\end{table}
%%%%%%%%%%%%%%%%%%%%%%%%%%%%%%%%%%%%%%%%%%%%%%%%%%%%%%%%%%%%%%%%%%%%%%%%%%%

Other measurements under consideration include \CP violation searches in modes with neutral pions.
Ongoing studies are targeted at establishing the efficiency and purity that can be achieved in these channels.
An area unique to LHCb is the search for \CP violation in charmed baryons which will play an important role in the upgrade era as well.

%%%%%%%%%%%%%%%%%%%%%%%%%%%%%%%%%%%%%%%%%%%%%%%%%%%%%%%%%%%%%%%%%%%%%%%%
\begin{figure}[htb]
\centering
\includegraphics[width=0.7\textwidth]{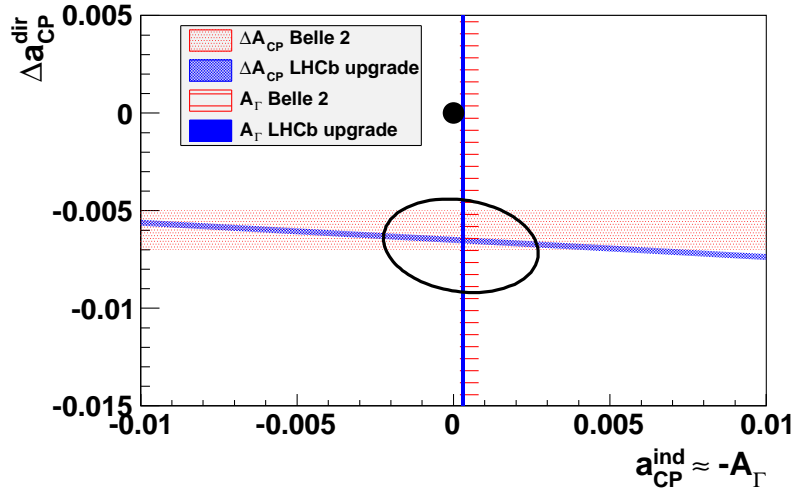}
\caption{Expected statistical sensitivities for \dacp and \agamma for Belle 2 and the LHCb upgrade. The central values are fixed to the current world average. The ellipse shows the current $1\sigma$ ellipse of the world average. The circle marks the no \CP-violation point with a radius of approximately $10^{-4}$.}
\label{fig:dacp_agamma}
\end{figure}
%%%%%%%%%%%%%%%%%%%%%%%%%%%%%%%%%%%%%%%%%%%%%%%%%%%%%%%%%%%%%%%%%%%%%%%%%%

A graphical example of the reach of the LHCb upgrade is shown in Figure~\ref{fig:dacp_agamma}.
It shows the comparison of the expected statistical sensitivities on \dacp and \agamma for Belle 2~\cite{Schwartz} and the LHCb upgrade.
Belle 2 will, together with continuously improving LHCb sensitivities, make a significant step forward in these measurements.
However, in order to positively identify indirect \CP violation beyond the standard model level the sensitivity of the LHCb upgrade will be required.

\section{Conclusions}
Measurements made with early LHC data have proven the feasibility of performing charm physics at a proton-proton collider.
In particular the high-precision \CP violation studies by LHCb show the way to a future of probing down to standard model precision levels with the LHCb upgrade.
Significant steps in precision are already expected with the current experiments until 2017.
Atlas and LHCb will further exploit the area of production and spectroscopy measurements.

The LHCb upgrade is mandatory to achieve the level of precision required to distinguish effects from standard model and beyond.
Through many complementary measurements of mixing and \CP violation observables the LHCb upgrade will pin down the underlying theory parameters.
Complementary to the area of \CP violation, the search for rare and forbidden decays will make a leap forward.

\bibliographystyle{utphys}       % APS-like style for physics
\bibliography{charm-lhc-upgrade}   % name your BibTeX data base
 
\end{document}